# Fragment-History Volumes

Francisco Inácio    Jan P. Springer

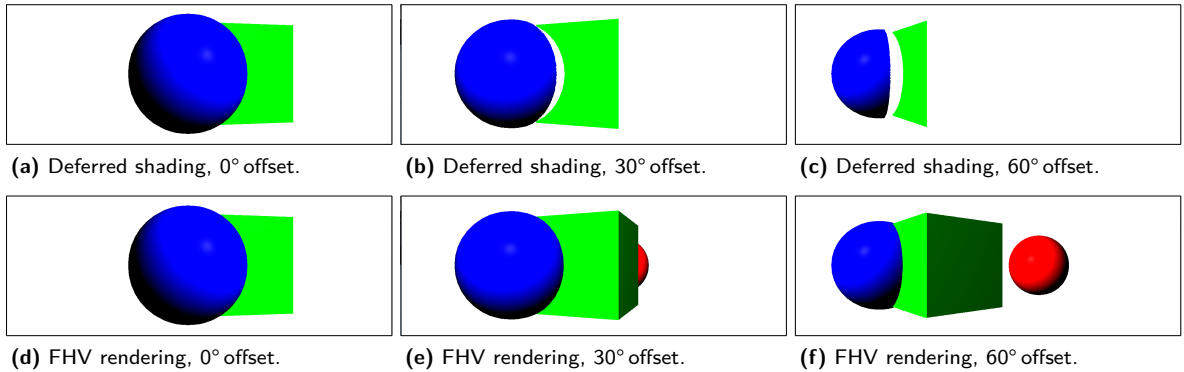

**(a)** Deferred shading, 0° offset.   **(b)** Deferred shading, 30° offset.   **(c)** Deferred shading, 60° offset.

**(d)** FHV rendering, 0° offset.   **(e)** FHV rendering, 30° offset.   **(f)** FHV rendering, 60° offset.

**Figure 1:** Screen captures (a) and (d) show a scene rendered using deferred shading and fragment-history volumes (FHV), respectively. (b) and (c) reuse the existing frame-buffer data to reconstruct for new viewpoints using deferred shading while (e) and (f) reconstruct the scene using FHVs for the same set of view points. While using our FHV technique in (e) and (f) allows for correctly reconstructing the scene for new viewpoints, deferred shading used in (b) and (c) exhibits artifacts and cannot reveal previously hidden objects.

**Abstract**: Hardware-based triangle rasterization is still the prevalent method for generating images at real-time interactive frame rates. With the availability of a programmable graphics pipeline a large variety of techniques are supported for evaluating lighting and material properties of fragments. However, these techniques are usually restricted to evaluating local lighting and material effects. In addition, view-point changes require the complete processing of scene data to generate appropriate images. Reusing already rendered data in the frame buffer for a given view point by warping for a new viewpoint increases navigation fidelity at the expense of introducing artifacts for fragments previously hidden from the viewer.

We present fragment-history volumes (FHV), a rendering technique based on a sparse, discretized representation of a 3d scene that emerges from recording all fragments that pass the rasterization stage in the graphics pipeline. These fragments are stored into per-pixel or per-octant lists for further processing; essentially creating an A-buffer. FHVs using per-octant fragment lists are view independent and allow fast resampling for image generation as well as for using more sophisticated approaches to evaluate material and lighting properties, eventually enabling global-illumination evaluation in the standard graphics pipeline available on current hardware.

We show how FHVs are stored on the GPU in several ways, how they are created, and how they can be used for image generation at high rates. We discuss results for different usage scenarios, variations of the technique, and some limitations.

**Keywords**: Rasterization, Real-Time Interactive Rendering, Scene Reconstruction, A-Buffer

# 1 Introduction

Maintaining a high steady frame rate is an important aspect in interactive real-time graphics, especially when used as a component in any of the artificial reality variants. It is mainly influenced by the number of objects and the number of lights to be processed for appearance evaluation in a given 3d scene. The upper-bound effort for rendering a scene is then defined by *the number of objects **times** the number of lights*, i.e. $\mathcal{O}(N_O \cdot N_L)$ meaning each object needs to be evaluated with each light. Deferred shading reduces this upper bound to *the number of objects **plus** the number of lights*, i.e. $\mathcal{O}(N_O + N_L)$, by separating the rendering process into two phases: a geometry-processing stage and a lighting-evaluation stage. The geometry-processing phase rasterizes all objects but only retains visible fragments in a G-buffer for the current viewpoint. The lighting-evaluation phase then only needs to process those surviving, i.e. visible, fragments to compute the final image (for the current viewpoint). Unfortunately, this approach not only trades computational effort for memory bandwidth but also requires the re-creation of the G-buffer every time the viewpoint, i.e. the viewer's camera(s), changes. Additionally, transparent objects cannot be encoded into a G-buffer and must be separately processed, usually in an additional render pass. Post-rendering 3d warping is one particular technique that allows to create images from existing G-buffer information for new viewpoints. However, this only works with sufficient fragment information. Objects not encoded into the G-buffer, because they were not visible from the original viewpoint, will create visual artifacts at discontinuities between (projected) objects (cf. top row in figure 1 for a visual example).

We developed fragment-history volumes (FHV) to support the creation of novel viewpoints from a discrete representation of the entire 3d scene using current graphics hardware and present an initial performance comparison. FHVs are based on the idea of the A-buffer and can be implemented on current graphics hardware as per-pixel linked lists of fragments (FHV$_{PPFL}$) or per-octant linked lists of fragments (FHV$_{POFL}$). Our prototype software uses shader-storage buffer objects (SSBO) in OpenGL 4.3+ but can also be implemented using analogous abstractions in DirectX 10+ or Vulkan. The capture of all fragments from a rasterized scene creates a record of all fragments for each pixel based on the order of fragments after rasterization. Hence the name fragment-history volumes. FHVs using per-pixel linked lists are biased towards the direction of the original viewpoint, i.e. the fragment density along the original view direction is much higher than for any orthogonal direction. Using simple point splatting to visualize the captured fragments exhibits artifacts for novel viewpoints. To alleviate this we show how to store the resulting fragments into an octree structure. Instead, of using per-pixel linked lists of fragments we use per-octant linked lists of fragments and rasterize each triangle to produce an optimum amount of fragments. The per-octant linked list of fragments is further developed into a per-octant array of fragments (FHV$_{POFA}$), which keeps fragments that belong to an octant in a contiguous range, potentially better supporting coherent memory-access patterns on current GPUs. Both per-octant variants create a view-independent representation of the original scene on the GPU. They also allow for spatial-index access into the discretized scene fragments with image-order techniques (e.g., ray casting) to create images for novel viewpoints at high frame rates. Additionally, this also enables transparency resolution, shadows, and indirect light evaluation on current graphics hardware.

Our main contributions are:

- We present fragment-history volumes, a view-independent representation of an A-buffer suitable for current GPU architectures.
- We show how hardware rasterization on the GPU can be employed to optimize fragment generation to create view-independent A-buffers by rasterizing triangles using each triangle's normal to build its tangent-space transform and storing the results in world-space coordinates.
- We show two distinct techniques for view-dependent reconstruction from fragment-history volumes. We evaluate the performance of our approaches with respect to deferred shading and discuss these results as well as their limitations.

To the best of our knowledge we are the first to propose a view-independent extension to the A-buffer idea as well as to show how fragment generation using a triangle's tangent-space transform allows for optimizing the memory requirements for view-independent A-buffers.

## 2 Related Work

Hardware implementations of the triangle rasterization pipeline are based on the idea of the Z-buffer [Catmull 1974]. An output-sized array of depth values is used to determine if a new fragment lies in front of the current fragment at the same location in the frame buffer and needs to replace the frame-buffer content to become visible. Essentially, this creates a *visible shell* of the scene for the current viewpoint after all geometry has been processed. Alpha compositing [Porter and Duff 1984] allows to resolve transparency by blending two fragments. However, once the scene is completely processed all information about which fragments made up the appearance of a pixel is simply lost. Changes in view direction as well as lighting require the complete reprocessing of the entire scene. Current graphics hardware, while based on the same principle(s), provides two additional properties. One is the ability to output more than a single four-component value from the fragment stage. The other is a transition from specialized processing hardware for programmable shader stages (e.g., vertex, fragment) to generalized processing cores [Hughes et al. 2013]. By supporting multiple output values from the fragment-shader stage a G-buffer [Saito and Takahashi 1990; Deering et al. 1988] is available that can hold intermediate computations per fragment (e.g., position, normal, diffuse color). This allows for separating the rendering process into a geometry-processing and a lighting-evaluation phase. The resulting technique, deferred shading [Hargreaves and Harris 2004], reduces the upper-bound effort for rendering a scene from *number of objects **times** number of lights* to *number of objects **plus** number of lights*. It also takes advantage of the generalized cores on current GPUs because the lighting pass only needs to evaluate at most the number of output pixels, which is achieved by submitting an output-resolution-sized quad from which the rasterizer creates exactly one fragment per pixel. The graphics hardware is then able to schedule all available cores for processing these fragments because no further vertex processing is required. Scene input still requires the sorting of geometry along the current view direction to achieve correct blending of fragments for transparency resolution in an additional render pass. Order-independent transparency support on graphics hardware became available with more flexible implementations of the raster pipeline by using depth peeling [Everitt 2001], *k*-buffer variants using multiple render targets [Bavoil et al. 2007], or bucket-sorted depth peeling [Liu et al. 2009]. These techniques either require multiple passes or restrict the number of fragments that can be resolved per pixel.

Yang et al. [2010] show how to create an A-buffer [Carpenter 1984] on the GPU and use it for order-independent transparency resolution. Their implementation is based on DirectX 10+ using struc-tured buffers for the pixel array and the linked list of fragments. Similar results can be achieved in OpenGL 4.3+/Vulkan using shader-storage buffer objects. Our FHV variant based on per-pixel linked lists of fragments (FHV$_{PPFL}$) is derived from those. S-buffer provides an alternative to a linked list of fragments by using variable contiguous regions for each pixel [Vasilakis and Fudos 2012]. Our per-octant array of fragments (FHV$_{POFA}$) follows a similar idea but stores fragments in an view-independent way. Transparency resolution requires the sorting of fragments per pixel along the current view direction. Lefebvre et al. [2013] use a hashing scheme that allows for efficiently sorting fragments on insertion into a fragment list. Knowles et al. [2014] discuss how sorting speed is impacted by local memory latency and occupancy, while Schollmeyer et al. [2015] provide a performance evaluation of different sorting strategies on GPUs. However, view-point changes still require the complete processing of the scene with any of those techniques.

Re-projection techniques (e.g., [Mark et al. 1997; Shade et al. 1998]) use already created shading information for some viewpoint and apply a warping transformation for a novel viewpoint. Because no information about discarded scene geometry is available this will create dis-occlusion errors if the two viewpoints diverge too much (q.v. figures 1a–1c). Reverse re-projection [Nehab et al. 2007] mitigates this problem by using a distance metric to decide if an existing result for some raster location can be used at its new location for a novel viewpoint. If not, the appearance will be recomputed and a shading cache updated. On average, for a small divergence of the viewpoint between frames, this will reduce the amount of fragments that require full appearance evaluation. However, because no spatial information about the scene is available, these techniques do not have the potential for applying indirect lighting evaluation. This is in contrast to our technique where the complete scene information is encoded into an octree structure [Meagher 1982], allowing spatial-index access, instead of an array of pixels or fragments. Additionally, dynamically updated lights will restrict the amount of shading information that can be stored in a shading cache.

Sparse voxel octrees [Crassin et al. 2009; Laine and Karras 2011] are memory-efficient spatial data structures that allow easy traversal for ray look-ups. They encode volumetric data or even triangle meshes as small voxel volumes called bricks, which are stored in texture memory. Bricks encoding the original data can be used to create coarser representations at lower levels, which is achieved by using mipmap filtering [Crassin et al. 2011]. These bricks are pointed to by octree nodes at the respective level in the octree. The octree is then used for ray traversal. Our approach similarly creates an octree but instead of pointing to bricks, our leaf nodes point to contiguous arrays of fragments that belong to the bounding volume of an octant. This dual structure, i.e. the octree as a spatial index structure plus a list or array of fragments, allows for both simply processing all the fragments with point-based rendering, i.e. ignoring the spatial indexing, as well as ray traversal. In [Crassin et al. 2011] a triangle-based scene is rasterized three times in order to capture all the necessary surface data and encode it as voxels. Our octree-based FHV approaches allow for rasterizing the triangle-based scene only once by maximizing the surface data captured with a normal-space projection per triangle. Crassin et al. [2011] use a $512^3$ octree setup and then rasterize at a resolution of $512 \times 512$, which means leaf-node rows match perfectly with raster or pixel locations. Our approach similarly uses a $512^3$ octree but employs a rasterization resolution of $1000 \times 1000$ or even higher. This allows us to store several fragments per octree node and thus increases the resolution of the scene representation. Hu et al. [2014] describe a ray-scene intersection technique based on rasterization on current GPUs. However, their method uses a uniform voxel grid in addition to an A-buffer, which in turn is based on [Yang et al. 2010], plus deferred shading for capturing surface appearance values. Additionally, the method is not view-point independent as well as seems to be at least one magnitude slower than our approach. Archer et al. [2018] describe a method that uses a compound deep image, which is created similarly to our FHV variant using a pre-allocated array of fragments (FHV$_{POFA}$), but uses virtual point lights for ray casting the final scene. However, their compound deep image is still a view-depended storage technique.

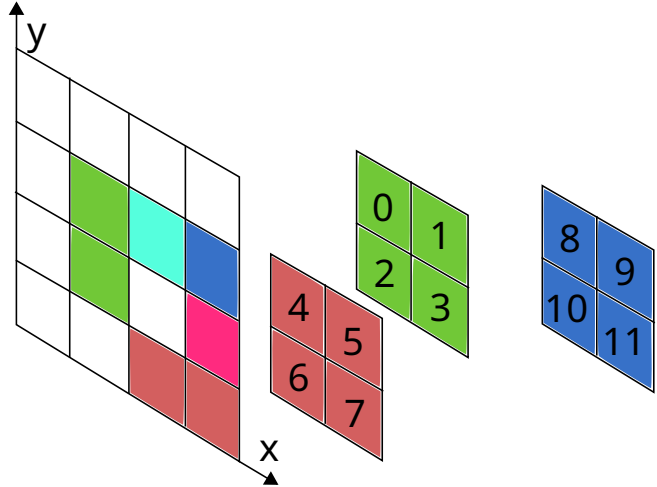

**Figure 2:** Example scene with three quads and the expected alpha composition in the display. The numbers in the quads indicate the (assumed) order of fragments produced by the rasterizer, i. e. first the green quad is rasterized, producing fragments 0, 1, 2, and 3, then the red quad is rasterized, producing fragments 4, 5, 6, and 7, and, finally, the blue quad is rasterized, producing fragments 8, 9, 10, and 11. However, the quads are ordered along the view axis from near to far as red, green, and blue. The pixel values in the screen are the result of compositing the fragments of the green quad with the fragments of the red quad and then with the fragments of the blue quad (assuming alpha values of $1/3$ for each of the quads).

# 3 Technique

Fragment-history volumes are a discrete representation of a 3d scene produced by the processing of geometry and associated attributes in a hardware-based feed-forward rasterization pipeline. There are several ways for storing the discrete information, i. e. fragments, and associated creation procedures, which are described in section 3.1 and 3.2, respectively. In addition, several techniques are available for reconstruction of novel viewpoints and are described in section 3.3.

## 3.1 FHV Storage

Storing multiple fragment values per pixel is currently not supported as a standard technique in the hardware graphics pipeline. Although recent graphics shader models (e. g., [Segal and Akeley 2014]) allow for more flexible customization of shader-pipeline stages the final frame-buffer setup is (usually) restricted to a four-component color image and a single-component depth-stencil image. Fragments passing the rasterization stage will be composited with the current frame-buffer content for their pixel location according to a user-defined blend function and replace that pixel value in the frame buffer. Fragments that do not pass the rasterization stage or the final z-test do not participate in this composition. In any case the information of what fragments would make up the discretized geometry (along the current view direction) is simply lost. In addition to the standard frame-buffer setup multiple render targets allow for storing up to eight four-component values into dedicated buffers plus a single-component depth-stencil buffer. While this increases the amount of information that can be retained after rasterization it still limits the number of fragments that could be stored per pixel position.

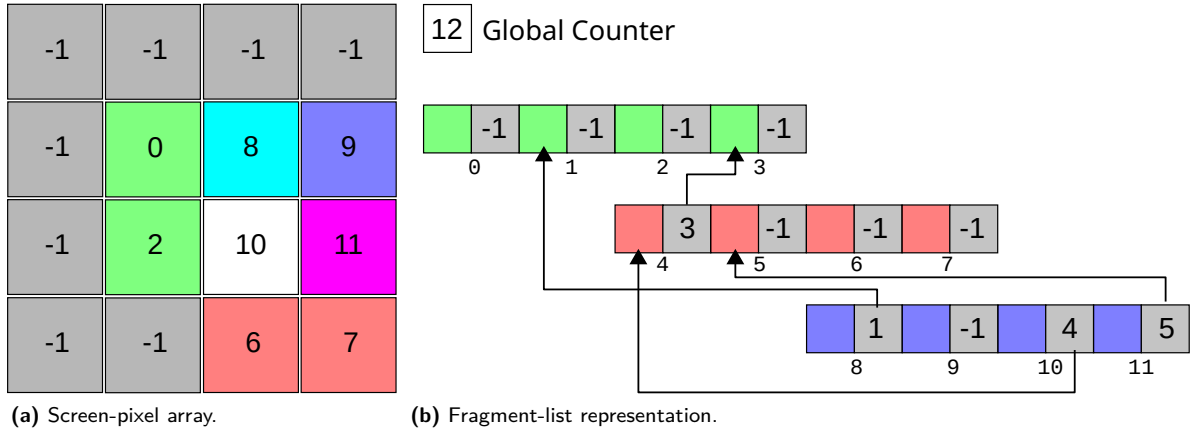

**(a)** Screen-pixel array.  **(b)** Fragment-list representation.

**Figure 3:** Per-pixel fragment lists using a simple storage-layout variant (FHV$_{PPFL}$) for captured fragments to build an A-buffer on current graphics hardware (based on the example scene in figure 2). (a) shows the array of pixel positions containing the index into the per-pixel fragment list where the last inserted fragment for that pixel position can be found (or -1 if no fragment was processed for that pixel location). (b) shows the global counter for the next free element (top) and the storage layout of the linked lists of fragments (bottom).

Figure 2 shows an example scene, which is used further in the text as well as in the following figures for describing the storage of fragments in a per-pixel as well as per-octant linked list or array of fragments. Note that in figure 2 the green quad is rasterized first, followed by the red and blue quads, while the order along the z direction, i.e. the current view direction, is red quad, green quad, and blue quad. The pixel values in the screen of figure 2 are the result of compositing the fragments of the green quad with the fragments of the red quad and then with the fragments of the blue quad under the assumption of fragments containing alpha values of 1/3 for each of the quads.

**Simple Layout** Current definitions of graphics-shader languages (e.g., [Kessenich 2014]) also allow for user-defined buffer setups. While these are still restricted to pre-allocated linear arrays they can be used to store almost any arbitrary user-defined data. FHVs use such buffers for two interconnected array structures: a *pixel array* and a *fragment list* (cf. figure 3). In the pixel array the last processed fragment for a pixel location is stored as an index into the global fragment list or -1 in case no fragment was (yet) recorded for that pixel position. A global counter is used to keep track of the next free index in the global fragment list. Atomic-increment operations are available to avoid inconsistent values while concurrently updating this global counter. In the global fragment list fragment data is stored together with the index to the previous fragment for the same pixel position or -1 if no such element is available.

While such structures (usually) are easy to realize in standard programming languages they are currently not available *as a primitive* on GPUs. The main reason for this is that they require dynamic memory allocation which is not (yet) supported on graphics hardware.[1] By using a dual structure of a pixel array storing indices to the start of a pixel position's fragment list and a global list of all recorded fragments it is possible to workaround this limitation, at least for the specialized case of FHVs. Both

---

[1] This is not completely correct. CUDA 6.0 and newer allows for dynamic memory allocation for user defined kernels. However, CUDA [Cook 2012] is a compute oriented API and vendor-specific graphics-firmware modes used by all graphics APIs are much more restrictive, especially with respect to memory control mechanisms, so as to guaranty certain performance specifications.

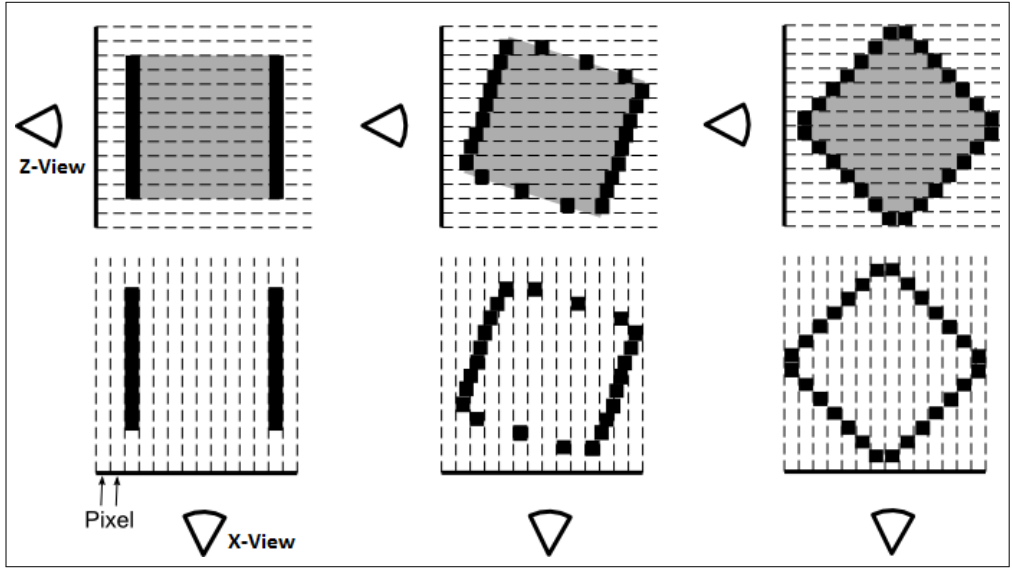

**(a)** Schematic View.

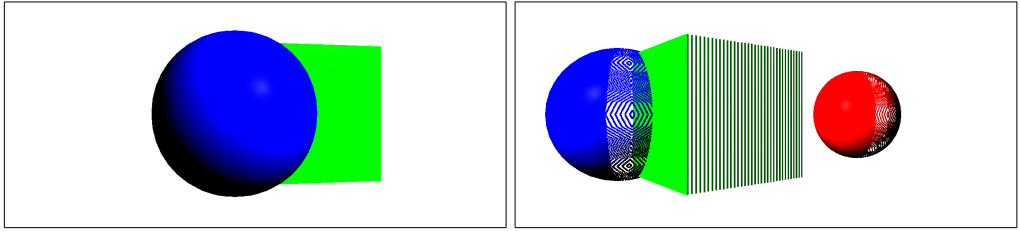

**(b)** X/Y direction.

**(c)** Z direction.

**Figure 4:** Fragment-distribution bias in an FHV$_{PPFL}$. (a) is a schematic view of the fragment density bias assuming fragments are created with the view direction along the z axis. (b) shows the fragment distribution for the original x and y directions. (c) shows the fragment distribution for the same global fragment list along the z direction with the FHV rotated by 60°.

arrays need to be pre-allocated. In the case of the pixel array the number of elements is simply the resolution of the rasterizer setup. The fragment list can be pre-allocated by multiplying the rasterizer resolution with some average depth complexity and the size of the actual fragment-data record to be stored; this is discussed in more detail in section 5. However, this size may be exceeded at run time and may require re-allocation, which is possible for the next frame and needs to be triggered from the controlling process on the CPU.

In order to minimize the memory requirements for a single fragment we only store its position and normal as well as its material and object id. The material id is an index into a global buffer holding all materials in a scene. The object id uniquely identifies the logical object the fragment belongs to. Using a 32-bit value for the index of the previous fragment this amounts to $3 \times 32$-bit for the fragment's position, $3 \times 32$-bit for the fragment's normal, and $3 \times 32$-bit for the material and object id as well as the index to the previous fragment. In total storing a fragment requires 36 bytes (or 48 bytes to adhere to four-component alignment requirements on current GPUs). We will refer to this simple per-pixel fragment-list layout as FHV$_{PPFL}$ for the remainder of the paper.

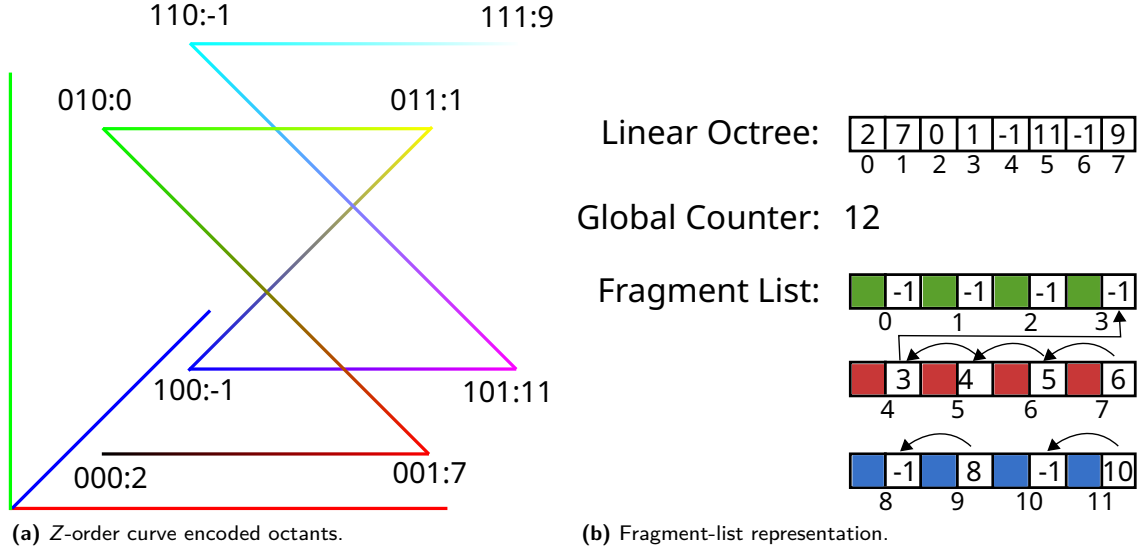

**(a)** *Z*-order curve encoded octants.      **(b)** Fragment-list representation.

**Figure 5:** Per-octant linked lists of fragments storage-layout variant (FHV$_{POFL}$) for captured fragments to build a view-independent A-buffer on current graphics hardware (based on the example scene in figure 2). (a) shows a *z*-order encoding where the first three digits are the *x*, *y*, and *z* components of a 3d location with the index of the last processed fragment appended after the colon. Main axes are color coded as well as the *z*-order curve, which shows the interpolation of eight corner values. (b) shows mapping of the Morton encoded spatial indices into a linear array (top) and the global list of fragments (bottom) similar to the simple FHV$_{PPFL}$ layout in figure 3b.

**Octree Layout** The simple layout FHV$_{PPFL}$ is biased toward the original view direction at the time of its creation. If the FHV is rotated around its *y* axis it can be seen that the fragment distribution along the original *z* axis is much less dense than for the original *x* and *y* directions (cf. figure 4). Creating an image for a new viewpoint different from the original viewpoint will then exhibit artifacts (cf. figure 4c). This can be avoided by creating the global fragment list from all three main directions. However, it will also require three pixel arrays, which must be created by rendering the scene three times and then (sequentially) queried when reconstructing images for novel viewpoints.

Our octree-based layout replaces the pixel array(s) of the FHV$_{PPFL}$ with an octree structure [Meagher 1982] where the fragment index into the global fragment list for the last processed fragment is stored in the leaf octants according to their spatial location. Normally, this would require a recursive data structure and dynamic memory allocation. However, an octree can also be implemented using a linear list and spatial indices processed using Morton code or *Z*-order curve encoding [Morton 1966] (cf. figure 5). A simplified view on Morton encoding would be similar to creating hash values from 3d spatial positions. To allow for efficient spatial look-up based on rays, including empty-space skipping optimization, lower-level octants are created, i.e. coarser levels in the octree, as linear arrays similar to the leaf octants but only storing occupancy information of their children octants. By including fragments transformed to world-space coordinates the octree-based layout creates a *view independent FHV*. We will refer to the octree-based per-octant linked list of fragments as FHV$_{POFL}$ for the remainder of this paper.

The global fragment list contains fragments in the order they were processed by the rasterizer and inserted by a fragment-shader program. A global counter indicating the next free fragment slot in the global list is required, which is atomically incremented for each fragment. We also developed a variant of the global fragment list that avoids this potential synchronization point. If the amount of fragments per

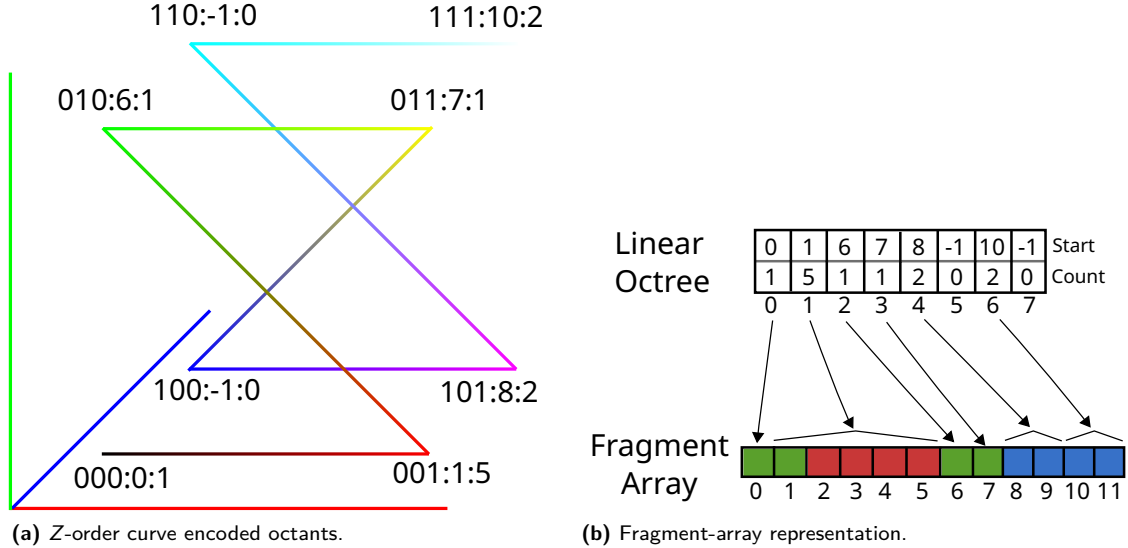

**(a)** Z-order curve encoded octants.  **(b)** Fragment-array representation.

**Figure 6:** Per-octant array of fragments storage-layout variant (FHV$_{POFA}$) for captured fragments to build a view-independent A-buffer on current graphics hardware (based on the example scene in figure 2). (a) shows a z-order encoding where the first three digits are the x, y, and z components of a 3d location with the index of the last processed fragment appended after the colon. Main axes are color coded as well as the z-order curve, which shows the interpolation of eight corner values. (b) shows mapping of the Morton encoded spatial indices into a linear array (top) and the global array of fragments (bottom). Note that the array is pre-allocated allowing for insertion of consecutive fragments into contiguous parts of the array.

leaf octant is a priori known then each octant can be assigned a fixed-sized part of the global fragment array where its octant-local fragments will be contiguously stored. While this avoids synchronizing with a single counter for all fragments to be processed it implies a two-pass setup approach. First the fragments per leaf octant are counted and recorded in the octant structure by rasterizing the scene. Each octant is assigned an index into the global fragment array based on the sum of all fragments from previous octants with respect to their Morton encoding. The scene is then rasterized again and incoming fragments are actually inserted into their pre-computed position in the global fragment array (cf. section 3.2). A schematic drawing for the simple case of a two-level octree is shown in figure 6. We will refer to the octree-based per-octant array of fragments as FHV$_{POFA}$ for the remainder of the paper.

## 3.2 FHV Creation

FHVs are created by rendering a scene and recording all fragments produced by the rasterizer. The vertex-shader program creates a standard homogeneous position for the rasterization as well as transforms the vertex' position and normal into world coordinates to be used for interpolation in the rasterizer between vertex positions of a triangle. Together with the fragment's world position, normal, as well as material and object id a fragment entry is created in the fragment-shader program and inserted into the global fragment list. The fragment-processing stage does not output any value into a frame buffer but simply discards the current fragment.

In our prototype implementation we render the entire scene using an orthographic projection to avoid unevenly distributed depth values for fragments. This will later help to simplify reconstruction for novel viewpoints. We also disable all visibility testing (e. g., frustum/back-face culling, scissoring, $z$-culling) to allow all fragments from the scene to be recorded. This will also record *invisible fragments* behind the viewer's current position. Doing so provides us with the opportunity to position the camera at the geometric center of the scene's bounding box.

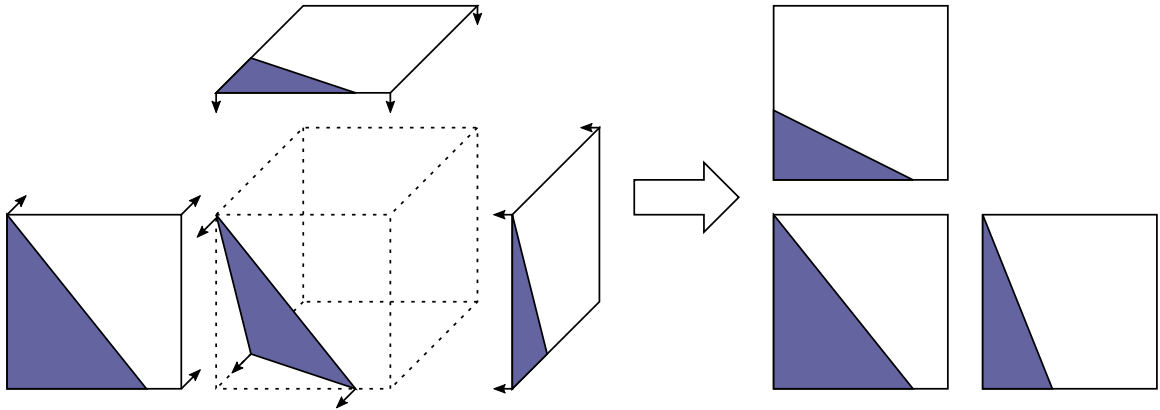

**Figure 7:** Cube rasterization. Rasterizing the scene from three main axes in order to capture all surface points, essentially creating three version of the same triangle from different viewpoints as well as duplicate surface data.

**FHV$_{PPFL}$**   In the fragment-shader program each incoming fragment's screen position in $x$ and $y$ is used to compute a unique index into an array of pixels for the FHV$_{PPFL}$. Fragment insertion into the global fragment list first obtains the next free index into the fragment list using an atomic operation incrementing a global fragment-index counter. If the pre-allocated size of the fragment list is still larger than that index the fragment content is written to that position in the global fragment list. To keep track of already recorded fragments the current fragment-list index for the current pixel location is stored as the index to the previous fragment in the new fragment's attribute record while the new fragment index is stored into the pixel array at the current pixel position (cf. figure 3). This task uses an atomic-exchange operation to avoid inconsistent index values in case of concurrent memory access for the current pixel position.

**FHV$_{POFL}$**   The fragment insertion into the global fragment list is the same for both FHV$_{PPFL}$ and FHV$_{POFL}$. However, while FHV$_{PPFL}$ simply uses the $x$ and $y$ fragment position to find the correct index in the pixel array, FHV$_{POFL}$ uses Morton encoding for the $x$, $y$, and $z$ fragment positions, which determines the index position into a linear array that represents the leaf nodes of an octree. For both the FHV$_{PPFL}$ and the FHV$_{POFL}$ this index position contains the last index into the the global list of fragments or -1 if no fragment was stored for that spatial index (cf. figure 5). In addition FHV$_{POFL}$ also updates parent octant's children occupancy, if necessary.

To avoid spatial bias of the stored fragments, FHV$_{POFL}$ ideally needs to *capture* the scene from the three main directions, essentially rendering the scene three times (cf. figure 7). Alternatively, this can be reduced to a single pass by transforming each vertex into world space in a vertex-shader program and processing the resulting triangle in a geometry-shader program. This geometry-shader program outputs the three projections of the triangle with respect to the $x$, $y$, and $z$ main directions. While this avoids the directional bias towards the current view direction it also creates an excessive amount of fragments as well as several fragments for the same spatial location at corners and edges shared by triangles. Our actual solution exploits the fact that the FHV$_{POFL}$ is view independent. This means that we are free to decide from which direction we want the (projected) triangle to be rasterized. Instead of creating three new triangles from the input triangle in the geometry-shader program it is sufficient to apply a view transform to that triangle that maximizes its projected size in the rasterizer. This transform is simply the tangent-space matrix of the triangle, i.e. it contains the tangent, bi-tangent, and normal of the triangle, which essentially rasterizes the triangle *as seen* from its surface normal (cf. figure 8). A performance comparison between these variants for creating an FHV<sub>POFL</sub> is provided in section 4.

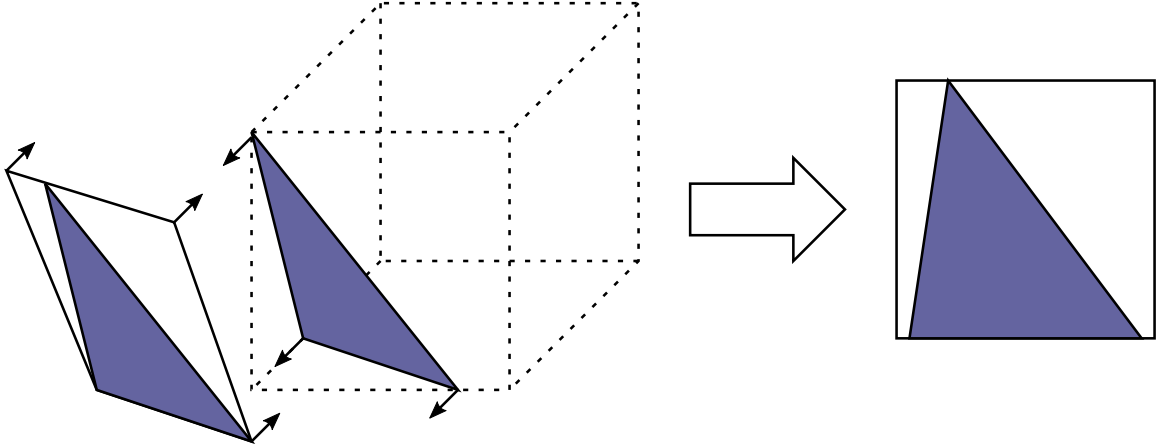

**Figure 8:** Tangent-space rasterization. Rasterizing the scene by using a projection plane parallel to each triangle's surface. Only one version of each triangle is produced.

**FHV<sub>POFA</sub>** The view-independent nature of FHV<sub>POFL</sub> allows for intersecting rays with the octree structure and to determine fragment ordering along that ray. However, the linked list of fragments contains fragments in the order of rasterization. Accessing the content of the fragment record (e.g., index of previous fragment) may create a memory request for data not local to a core within a thread group. FHV<sub>POFA</sub> avoids this by contiguously storing fragments that belong to an octant. This is achieved by first processing the complete scene and storing fragment counts per octant. This information allows for correct pre-allocation of the global array as well as the assignment of per-octant offsets into the global array. The per-octant counters are reset to zero and reused in the FHV creation pass as per-octant counters to the next free fragment in the per-octant fragment array starting at the previously determined offset. The final values per octant for the offset and the fragment counter can then be used during image generation for novel viewpoints to limit access to the correct range into the global fragment array.

## 3.3 Image Generation for Novel Viewpoints

The global list of fragments can be processed using point-based rendering [Pfister et al. 2000] on the GPU. A vertex-shader program processes each fragment according to the current view-projection setup. It is sufficient to submit the amount of point indices from the controlling process on the CPU, then each vertex-shader program invocation can access a fragment using a unique id (e.g., `gl_PrimitiveID` in GLSL). Transformed points are further processed by a geometry-shader program. Based on the size of a pixel (in world space) each point is transformed into a quad of two triangles that will fill (at least) one pixel. Finally, these triangles are rasterized and create fragments that are evaluated with respect to appearance properties in a fragment-shader program. Figures 9a–9d show screen captures for images generated from an FHV$_{\text{PPFL}}$ using simple point-based rendering. Figures 9e–9h show screen captures for images generated from an FHV$_{\text{POFL}}$ also using a simple point-based rendering technique.

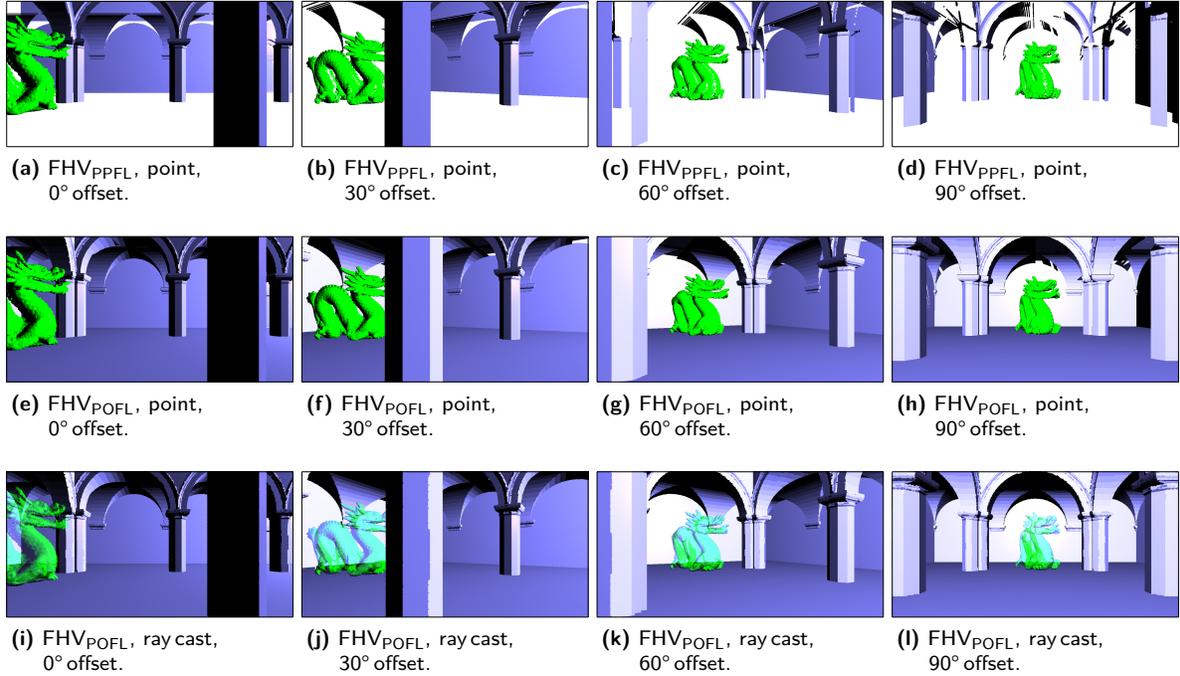

**Figure 9:** (a)–(d) Screen captures of a scene showing an FHV$_{\text{PPFL}}$ reconstructed with point-based rendering; note: visible artifacts, no transparency. (e)–(h) Screen captures of a scene showing an FHV$_{\text{POFL}}$ reconstructed with point-based rendering; note: *no visible artifacts*, no transparency. (i)–(l) Screen captures of a scene showing an FHV$_{\text{POFL}}$ reconstructed with ray casting; note: *no visible artifacts*, *resolved transparency*.

The rasterization setup is not required to be as restricted as for the FHV-creation process and may invoke standard optimizations to minimize the amount of generated fragments. However, this will usually lead to appearance similar to deferred shading and may be classified as a deferred-shading emulation when using only the fragment nearest to the current viewpoint. It will also exhibit artifacts for FHV$_{\text{PPFL}}$, especially if viewed from orthogonal directions to the original view direction, which can be controlled to a some extend by adapting the size of the quad or by using geometric deformation techniques. Geometric deformation is easier in our case because, in contrast to rendering point clouds, the fragments in the global fragment list already contain normals. Using connectivity information between points in the geometry-shader program allows for introducing attributes to be interpolated between those points without the need to establish (partial) surface estimation. Evaluating the global list of fragments using a point-based rendering approach essentially processes the entire scene twice. The first time when the FHV is created and the second time when all fragments are processed again by the vertex, geometry, and rasterization stages of the hardware pipeline.

The excessive geometry processing when using a point-based rendering approach for the global list of fragments can be avoided by adopting an evaluation approach based on ray casting. Submitting a full-screen quad to the graphics pipeline will create exactly one fragment-shader program invocation per

|        |       | Simple Scene |     | Complex Scene |      | Simple Scene |     | Complex Scene |     |
|--------|-------|------|-----|-------|------|------|-----|------|-----|
|        | Frame # | 0 | 1+ | 0 | 1+ | 0 | 1+ | 0 | 1+ |
| DS     | Geom  | 0.2  | 0.2 | 19.5  | 19.5 | 0.3  | 0.3 | 5.6  | 5.6 |
|        | Eval  | 0.3  | 0.3 | 0.2   | 0.2  | 0.3  | 0.3 | 0.3  | 0.3 |
|        | Total | 0.5  | 0.5 | 19.7  | 19.7 | 0.6  | 0.6 | 5.9  | 5.9 |
| FHV$_{PPFL}$ | Geom  | 10.3 | 0.0 | 102.2 | 0.0  | 1.3  | 0.0 | 6.0  | 0.1 |
|        | Eval  | 1.2  | 1.2 | 1.3   | 1.3  | 2.6  | 0.8 | 2.9  | 2.6 |
|        | Total | 11.5 | 1.2 | 103.5 | 1.3  | 3.9  | 0.8 | 8.9  | 2.7 |
| FHV$_{POFL}$ FHV$_{POFA}$ | Geom  | 7.6  | 0.0 | 99.5  | 0.0  | 1.6  | 0.1 | 13.9 | 0.0 |
|        | Eval  | 1.2  | 1.2 | 1.4   | 1.4  | 1.8  | 1.6 | 3.9  | 3.8 |
|        | Total | 8.8  | 1.2 | 100.9 | 1.4  | 3.4  | 1.7 | 17.8 | 3.8 |

(a) NVIDIA Geforce GTX 980M  (b) NVIDIA GeForce RTX 2070 Super

**Table 1:** Performance comparison between deferred shading (DS), simple FHV (FHV$_{PPFL}$), and octree-based FHV (FHV$_{POFL}$, FHV$_{POFA}$). For each technique the (average) time in milliseconds is given for the geometry pass, the evaluation pass using point-based rendering, and the sum of both with respect to a simple and a complex scene as shown in figure 10. Note that the excessive times in the geometry pass for the FHV techniques are only necessary for the first frame, which is indicated as frame #0. Frames rendered afterwards are indicated by the column labeled frame #1+. Values shown in tables (a) and (b) were measured with an NVIDIA Geforce GTX 980M card using driver version 357.52 on Windows 8.1 and an NVIDIA GeForce RTX 2070 Super card using driver version 436.30 on Windows 10, respectively.

output pixel. Within each fragment-shader program a ray is constructed from the current view position through the pixel's location in world coordinates. This ray can then be intersected with an FHV$_{POFL}$ or FHV$_{POFA}$ to determine all fragments that lie along its path (cf. figures 9i–9l). The intersection test is started with the lowest-level octant for the whole scene. If any of its children contains fragments, those are recursivley visited until the process arrives at a leaf octant. Visiting children octants is determined by distance to the ray origin, i.e. the current view point. In an intersecting leaf octant the fragments are tested and if any are intersected by the ray, they are sorted into a local fragment list according to their distance with respect to the ray origin. Transparency information is accumulated per octant and used as a threshold to avoid unnecessary work. Fragments only need to be sorted within their octant and per-octant fragment arrays can simply be appended because octants are already evaluated from nearest to farthest.

## 4 Results

We evaluate the performance of our FHV technique with respect to deferred shading as well as with respect to each other in terms of run-time performance and memory consumption. All reported values were measured using an NVIDIA Geforce GTX 980M graphics card with driver version 357.52 on Windows 8.1 as well as an NVIDIA Geforce RTX 2070 Super card with driver version 436.30 on Windows 10.

We compare the performance of the geometry-processing stage for deferred shading (DS), FHVs using a per-pixel linked list of fragments (FHV$_{PPFL}$), and FHVs using an octree representation of fragments (FHV$_{POFL}$, FHV$_{POFA}$) in table 1. We used two different scenes for our tests: a simple low depth-

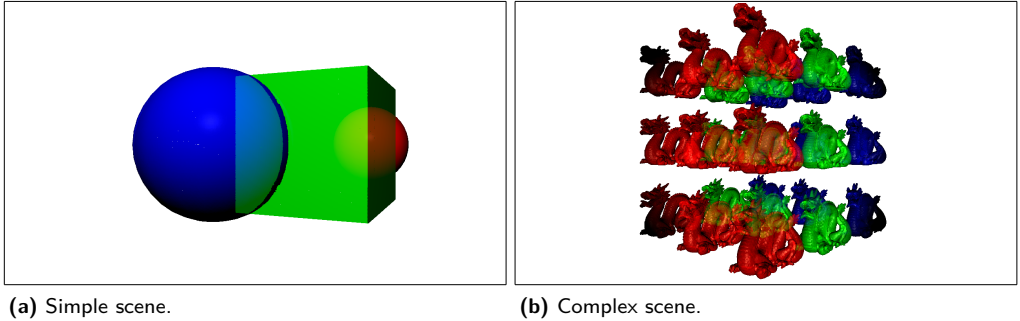

**(a)** Simple scene.      **(b)** Complex scene.

**Figure 10:** Screen captures of a simple and a complex scene used for obtaining performance values. Simple scene in (a) contains 1,532 triangles while the complex scene in (b) contains 23,525,262 triangles.

complexity scene and a scene with high polygon count as well as high depth complexity (cf. figure 10). Our tests using an NVIDIA Geforce GTX 980M setup indicate that, for DS with the simple scene, the geometry pass took 0.2 ms and the lighting-evaluation pass took 0.3 ms on average. The correspondent values for $FHV_{PPFL}$ and $FHV_{POFL}/FHV_{POFA}$ were 10.3/1.2 ms and 7.6/1.2 ms, respectively. For the complex scene DS took 19.5 ms for the geometry pass and 0.2 ms for the lighting-evaluation pass. The correspondent values for $FHV_{PPFL}$ and $FHV_{POFL}/FHV_{POFA}$ were 102.2/1.3 ms and 99.5/1.4 ms, respectively. Using an NVIDIA GeForce RTX 2070 Super setup the DS geometry pass took 0.3 ms and the lighting-evaluation pass took 0.3 ms on average. The correspondent values for $FHV_{PPFL}$ and $FHV_{POFL}/FHV_{POFA}$ were 1.3/2.6 ms and 1.6/1.8 ms, respectively. For the complex scene using an NVIDIA GeForce RTX 2070 Super setup DS took 5.6 ms for the geometry pass and 0.3 ms for the lighting-evaluation pass. The correspondent values for $FHV_{PPFL}$ and $FHV_{POFL}/FHV_{POFA}$ were 6.0/2.9 ms and 13.9/3.9 ms, respectively.

For static viewpoints on NVIDIA Geforce GTX 980M deferred shading clearly outperforms our FHV technique, i.e. the DS lighting-evaluation pass is, on average, ten times faster than either that of $FHV_{PPFL}$, $FHV_{POFL}$, or $FHV_{POFA}$. However, in the case of a (constantly) changing viewpoint DS needs to execute its geometry pass each frame while our FHV technique simply reuses the view-independent spatial representation of the scene. For the simple scene DS is still faster due to less overhead by either storing a fragment's data or simply discarding it after the rasterization stage. However, for complex scenes the DS's geometry pass becomes the bottleneck. In contrast, our FHV technique only needs to execute the lighting-evaluation pass. Frame times are then: DS 19.5 ms + 0.2 ms vs. $FHV_{POFL}/FHV_{POFA}$ 1.4 ms using an NVIDIA Geforce GTX 980M setup and DS 5.6 ms + 0.3 ms vs. $FHV_{POFL}/FHV_{POFA}$ 3.8 ms using an NVIDIA GeForce RTX 2070 Super setup.

Using newer hardware such as an NVIDIA GeForce RTX 2070 Super the picture changes to a more balanced run-time performance between DS and the simple $FHV_{PPFL}$ techniques with a slight advantage of $FHV_{PPFL}$ after the first frame. However, the $FHV_{POFL}/FHV_{POFA}$ techniques, which allow generation of novel viewpoints, are still approximately three times slower in their first frame but absolutely just miss the 16.6 ms mark for interactive real-time performance at 60 Hz.

Table 1 indicates an overhead for the initial creation of an FHV compared to DS. Table 2 compares the fragment creation for an FHV using four different methods: a single pass with the current view transform (1-Pass/View), three passes each for the x, y, and z main directions (3-Pass/1-Way), a single pass producing three triangles for each of the main directions in a geometry-shader program (1-Pass/3-Way), and a single pass using the tangent-space matrix for each processed triangle as the view transform (1-Pass/Normal) (q.v. section 3.2). All four methods have nearly the same CPU overhead for both

|          |       | 1-Pass / View | 3-Pass / 1-Way | 1-Pass / 3-Way | 1-Pass / Normal | 1-Pass / View | 3-Pass / 1-Way | 1-Pass / 3-Way | 1-Pass / Normal |
|----------|-------|---------------|----------------|----------------|-----------------|---------------|----------------|----------------|-----------------|
| Simple   | CPU   | 34.9          | 35.0           | 35.6           | 35.0            | 1.8           | 2.5            | 3.5            | 1.7             |
|          | GPU   | 10.4          | 11.2           | 11.2           | 10.8            | 1.0           | 1.2            | 1.1            | 1.0             |
|          | Calls | 3             | 9              | 3              | 3               | 3             | 9              | 3              | 3               |
|          | Frags | 231,096       | 678,768        | 678,768        | 514,479         | 231,096       | 678,768        | 678,768        | 514,479         |
| Complex  | CPU   | 35.0          | 35.6           | 35.2           | 35.0            | 2.1           | 2.3            | 3.7            | 3.8             |
|          | GPU   | 65.9          | 177.8          | 115.0          | 69.0            | 15.4          | 41.5           | 30.7           | 15.2            |
|          | Calls | 27            | 81             | 27             | 27              | 27            | 81             | 27             | 27              |
|          | Frags | 760,698       | 2,526,552      | 2,526,498      | 1,685,548       | 760,698       | 2,526,552      | 2,526,498      | 1,685,548       |

**(a)** NVIDIA Geforce GTX 980M    **(b)** NVIDIA GeForce RTX 2070 Super

**Table 2:** Performance comparison of the FHV creation process for the simple and complex scenes shown in figure 10. The average CPU and GPU time in milliseconds is given as well as the number of draw calls and the number of fragments created for: a standard single pass using the current viewpoint (1-Pass/View), three separate passes for $x$, $y$, and $z$ directions (3-Pass/1-Way), a single pass using a geometry-shader program to output triangles projected to $x$, $y$, and $z$ directions, respectively (1-Pass/3-Way), and a single pass using a triangle's normal as view direction (1-Pass/Normal). Values shown in tables (a) and (b) were measured with an NVIDIA Geforce GTX 980M card using driver version 357.52 on Windows 8.1 and an NVIDIA GeForce RTX 2070 Super card using driver version 436.30 on Windows 10, respectively.

hardware platforms as indicated in tables 2a and 2b. 1-Pass/View can be thought of as a baseline and exhibits the lowest execution times on the GPU. However, because an arbitrary view transform is used the fragments will be clustered in the plane orthogonal to that view direction and more sparsely along the view direction. The 3-Pass/1-Way method exhibits an increase in fragments but also nearly a tripling in GPU-processing time. This is somewhat alleviated with the 1-Pass/3-Way method, which reduces the GPU-processing time to approximately 66 % of the 3-Pass/1-Way method. The main difference between the 3-Pass/1-Way method and the 1-Pass/3-Way method is that the latter does not repeat the input assembly step, exhibiting a higher computational load on the GPU. 1-Pass/3-Way also needs less draw calls than 3-Pass/1-Way, which indicates that input assembly and vertex setup play a significant role for rendering performance on GPUs. Finally, the 1-Pass/Normal method is nearly as fast as the baseline case of 1-Pass/View while also creating less fragments than the 3-Pass/1-Way and 1-Pass/3-Way methods but more than the 1-Pass/View method. 1-Pass/Normal provides the optimum amount of fragments at nearly the same performance as the directionally biased 1-Pass/View. However, 1-Pass/Normal is only applicable to the view-independent fragment storage of FHV$_{POFL}$ and FHV$_{POFA}$. Additionally, run-time performance values in table 2 for the two hardware platforms indicate nearly a ten-fold performance increase of the NVIDIA GeForce RTX 2070 Super over the NVIDIA Geforce GTX 980M but the performance relationship between the four methods tested remains unchanged.

Table 3 compares the time for the lighting and material evaluation pass between DS, FHV$_{PPFL}$ using point-based rendering, and FHV$_{POFA}$ using ray casting. For FHV$_{POFA}$ timings were gathered for simple ray casting (R), ray casting with transparency resolution (R/T), and ray casting with transparency resolution as well as shadow computation (R/T/S). While FHV$_{PPFL}$ using point-based rendering is still faster than DS for the complex scene, FHV$_{POFA}$ using ray casting for lighting and material evaluation exhibits a clear performance hit in all use cases. However, it is worth noting that while the performance increase between the NVIDIA Geforce GTX 980M and the NVIDIA GeForce RTX 2070 Super holds as well, the performance gain of the NVIDIA GeForce RTX 2070 Super setup for the complex scene for simple ray casting is approximately ten times, for the ray casting with transparency resolution approximately

|  |  | DS | FHV$_{PPFL}$ | FHV$_{POFA}$ | | | DS | FHV$_{PPFL}$ | FHV$_{POFA}$ | | |
|---|---|---|---|---|---|---|---|---|---|---|---|
|  |  |  |  | R | R/T | R/T/S |  |  | R | R/T | R/T/S |
| Simple | Solid | 0.3 | 1.2 | 26.5 | 36.2 | 74.9 | 0.3 | 0.8 | 2.4 | 2.8 | 17.8 |
| Simple | Transp | — | — | 27.3 | 100.4 | 155.2 | — | — | 4.2 | 4.6 | 32.2 |
| Cmplx | Solid | 19.7 | 2.5 | 48.6 | 83.1 | 164.9 | 6.0 | 3.7 | 5.5 | 5.7 | 48.8 |
| Cmplx | Transp | — | — | 47.0 | 211.2 | 408.2 | — | — | 5.6 | 10.8 | 115.4 |

**(a)** NVIDIA Geforce GTX 980M        **(b)** NVIDIA GeForce RTX 2070 Super

**Table 3:** Performance comparison of lighting and material evaluation for deferred shading (DS), simple FHV (FHV$_{PPFL}$) using point-based rendering, and octree-based FHV (FHV$_{POFA}$) using ray casting. For DS and FHV$_{PPFL}$ no transparent objects were included. For FHV$_{POFA}$ simple ray casting (R), ray casting with transparency resolution (R/T), and ray casting with transparency resolution as well as shadow computation (R/T/S) were evaluated. All times are provided in milliseconds. Values shown in tables (a) and (b) were measured with an NVIDIA Geforce GTX 980M card using driver version 357.52 on Windows 8.1 and an NVIDIA GeForce RTX 2070 Super card using driver version 436.30 on Windows 10, respectively.

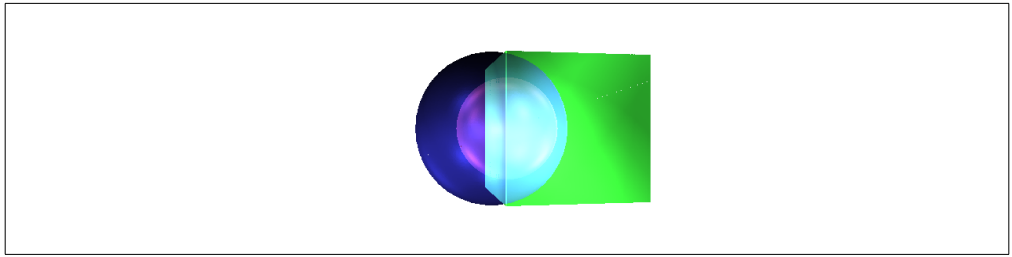

**Figure 11:** Simple scene rendered using ray casting with transparency evaluation.

twenty times and for ray casting with transparency resolution plus shadow computation only approximately four times. This indicates a non-linear relationship in performance behavior for these more sophisticated evaluation strategies and is further discussed in section 5. Figures 11, 12, and 13 show example screen captures of the simple scene, the hairball scene, and the Sponza, respectively, rendered using ray casting with transparency and shadow evaluation.

Table 4 compares the memory requirements for deferred shading and our FHV technique. Deferred shading solely depends on the output resolution while FHV depends on the *input resolution* of the rasterization stage. FHV$_{PPFL}$ and FHV$_{POFL}$ both over-allocate the global fragment list. This can be avoided by using a similar strategy as the S-buffer technique [Vasilakis and Fudos 2012] for FHV$_{PPFL}$ or replacing FHV$_{POFL}$ with FHV$_{POFA}$. The numbers shown in table 4 include a complete octree setup at various levels plus the global fragment list. For FHV$_{POFL}$ each octant requires $1 \times 4$ bytes plus the per-octant linked list of fragments. For FHV$_{POFA}$ each octant requires $2 \times 4$ bytes plus the per-octant fragment array. Besides the reduced storage requirements per fragment for FHV$_{POFA}$, because no index to the previous fragment needs to be stored, FHV$_{POFA}$ achieves a reduction by approximately an order of magnitude through the initial fragment counting step, which allows for exactly allocating the necessary amount of fragments. In contrast FHV$_{POFL}$, in our current prototype, uses an arbitrary estimation for the average depth complexity, which leads to over-allocation. This over-allocation for FHV$_{POFL}$ can be reduced by also including a fragment estimation step similar to FHV$_{POFA}$.

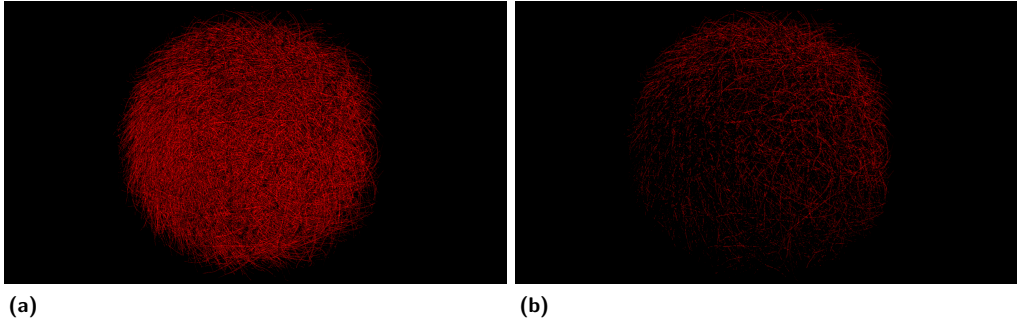

**(a)** **(b)**

**Figure 12:** Hairball model stored in an octree-based FHV (FHV$_{POFA}$) and rendered with ray casting. (a) shows transparency resolution but no shadows and (b) shows transparency resolution and shadows.

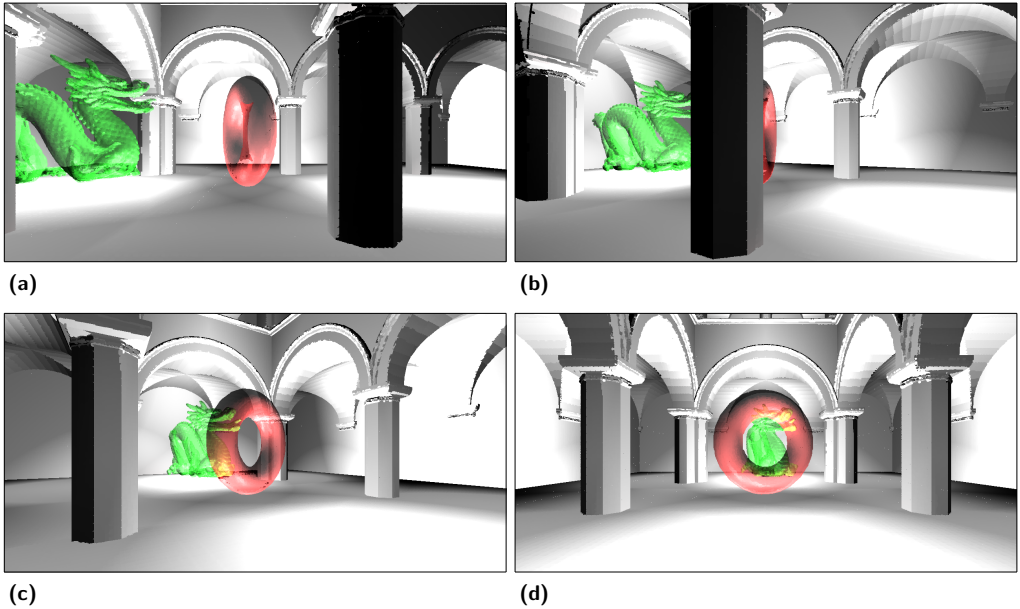

**(a)** **(b)**

**(c)** **(d)**

**Figure 13:** Sponza scene rendered from different angles using ray casting with transparency and shadow evaluation.

|  | Levels | Resolution | | Scene | |
|---|---|---|---|---|---|
|  |  | Input | Output | Simple | Complex |
| DS | — | — | $1280 \times 720$ | 28 MB | 28 MB |
| FHV$_{PPFL}$ | — | $1000 \times 1000$ | — | 462 MB | 462 MB |
| FHV$_{POFL}$ | 6 | $1000 \times 1000$ | — | 459 MB | 459 MB |
|  | 7 | $1000 \times 1000$ | — | 467 MB | 467 MB |
|  | 8 | $1000 \times 1000$ | — | 531 MB | 531 MB |
| FHV$_{POFA}$ | 6 | $1000 \times 1000$ | — | 18 MB | 54 MB |
|  | 7 | $1000 \times 1000$ | — | 34 MB | 70 MB |
|  | 8 | $1000 \times 1000$ | — | 162 MB | 198 MB |

**Table 4:** Memory-usage comparison between deferred shading (DS), FHV$_{PPFL}$, FHV$_{POFL}$, and FHV$_{POFA}$. For each technique the amount of memory is given with respect to the simple and complex scenes shown in figure 10.

# 5 Discussion

Our FHV technique is an evolution of earlier hardware-based A-buffer techniques. FHV$_{PPFL}$ is a simple and easy to implement technique but does not provide a view-independent storage format, which would be necessary for the creation of novel viewpoints without artifacts. This can only be achieved using a view-independent spatial data structure such as an octree as exemplified by FHV$_{POFL}$ and FHV$_{POFA}$.

The octree-based FHV methods are view-independent representations of a scene and allow for image generation for more than one novel viewpoint. This property can be exploited to generate stereo images for existing and emerging consumer head-mounted display devices, which separate the left- and right-eye view on a single display panel. In addition the same technique is extendable to multi-viewpoint setups for auto-stereoscopic devices. The limiting factor here will be the amount of time spent for generating an image for a single view multiplied by the number of required views. Each view would create a view-port-sized quad producing fragments used for lighting and material evaluation. However, because not all fragments will have the same work load submitting several view-port-sized quads along with the necessary attributes will (possibly) create a better computational occupancy of GPU cores and reduce the time to complete a frame of multiple view-point images.

We currently use a rather simple fragment-evaluation technique to determine pixel appearance. Essentially, this uses a local illumination model per fragment [Phong 1975; Blinn 1977] and accumulation of the fragment appearance to determine pixel appearance. However, because we have geometric information from all fragments from the scene, including which *logical object* a fragment belongs to, determining ambient occlusion or indirect lighting is feasible. By determining nearest fragments from the same or different objects additional rays can be created from the currently processed fragment location and evaluated for occlusion or radiance transfer. This will require a much more sophisticated model for describing material properties but would open the way to physically-based rendering techniques.

Fragment-history volumes provide a discretized and, in the case of octree-based FHVs, view-independent representation of an entire scene. However, the image generation for novel view points resamples this discretized representation at an arbitrary sampling rate determined by the output resolution of the resulting image. This will lead to artifacts in the final image depending on the *input resolution* with which the scene was discretized (q.v. table 4). If this resolution is too low resampling will create images with rather blocky appearance. On the other hand, if the input resolution is too high there might not be enough memory available on the GPU to contain all fragments. Additionally, the optimal input resolution is also connected to the output resolution of the final image as well as the projection setup for its viewport. This is similar to sampling problems for shadow-mapping techniques but in this case the resampling needs to process a 3d volume rather than a depth image. In our experiments we found that an input resolution of $1000 \times 1000$ is sufficient and probably already too much considering that this resolution will be used to rasterize each triangle from its normal direction. For other work loads, such as collision detection, much lower resolution rasterizer setups might be completely sufficient.

Directional bias of fragments in an FHV clearly affects reconstruction and final image fidelity. Rendering a scene three times, each for one of the main directions, incurs an additional overhead of an order of magnitude compared to deferred shading. However, for octree-based FHVs this is only true for the very first frame, which initializes the FHV structure. In addition for octree-based FHVs it is possible to optimize the amount of fragments by rasterizing each triangle according to its tangent-space transform in a single pass. As we have shown in our results in section 4, rasterizing the scene three times can be up to 250 % as costly as our tangent-space transform approach, which also explains the high creation cost of 280 ms in [Crassin et al. 2011]. In all of the following frames the view-independent information in the FHV can be used for image generation for novel viewpoints with much lower processing time using point-based rendering than deferred shading for scenes with high depth complexity (cf. table 1). We have also shown that the amount of draw calls, which requires input assembly and vertex-program setup, even if no CPU-to-GPU data transfer is necessary, plays a significant role for the time required to process a scene and to create an FHV representation from it (cf. table 2).

High performance in a fragment-shader program is tightly coupled to homogeneous execution of the same or similar steps for fragment-shader program invocations in a thread group on the GPU. This works very well for deferred shading where only a final fragment needs to be evaluated. $\text{FHV}_{\text{POFL}}$ and $\text{FHV}_{\text{POFA}}$ on the other hand using ray casting, even in our simple setup, evaluate all fragments contributing to a pixel by accumulating the fragments' appearance information. This not only means repeating the single evaluation of a fragment similar to deferred shading but also that the actual evaluation per fragment may diverge in terms of data each fragment requires and how this evaluation will proceed. This is because of how adjacent rays in the same thread group may reflect into different spatial directions and require data from different octants. Essentially, the thread group is then bound to wait for the slowest fragment-shader program. A similar performance argument holds also for the case that a single ray may split into multiple rays when using particular appearance evaluation models (e.g., reflection, refraction, multi-scattering). Archer et al. [2018] essentially point out the same, i.e. that coherently laid out fragment data, while incurring a slight run-time penalty in the setup process, will provide faster ray-casting performance because adjacently stored fragment data allows for coherent memory access. Additionally, Setaluri et al. [2014] observe that Morton encoding is akin to a locality-promoting map that ensures geometrically proximate entries will, with high probability, be stored in nearby memory locations. Consequently, the occupancy pattern of the $\text{FHV}_{\text{POFA}}$ fragment array as well, to a lesser extent, the $\text{FHV}_{\text{POFL}}$ fragment list will exhibit a high spatial coherency because the corresponding data entries will be highly clustered in memory. While this will help for ray-casting performance, more sophisticated techniques will still incur a performance hit when rays diverge and non-local data needs to be transferred for a single core in a thread group.

In its current form FHVs are pre-dominantly useful to capture large and detailed but static environments. However, the very fast appearance evaluation using point-based rendering allows for FHVs to be used as the base for rendering environments with dynamic elements. This can be as simple as rendering dynamic objects after creating an image for a novel viewpoint similar to rendering translucent objects after a deferred-shading pass. In addition, FHVs also allow for collision detection between dynamic objects and the static environment without the need for proxy representations of the static scene elements. Such a use case conceivably may not require the high input resolution for the rasterization step and will certainly benefit from the logical attributes available at the fragment level (e.g., object id, material id). As mentioned before it may use a lower input resolution or appropriate octree levels in an $\text{FHV}_{\text{POFL}}$/$\text{FHV}_{\text{POFA}}$. We also experimented with creating several FHVs according to update frequencies of dynamic objects. While this increases the overall memory requirements, each of these FHVs will on average contain a smaller amount of fragments, which implies a faster re-creation time as well as the ability to incrementally update parts of the scene for dynamic objects with a similar or the same update frequency. Updating these FHVs in an asynchronous way would provide a solution to the navigational aspect of multi-frame rate rendering [Springer et al. 2008].

# 6 Conclusions

We presented fragment-history volumes, a view-independent extension of the A-buffer realized on current graphics hardware that allows to generate images for (multiple) novel viewpoints. Starting with a per-pixel linked list of fragments (FHV$_{PPFL}$) we described how to store and create a view-independent discretized representation of a 3d scene using an octree for spatial classification of all fragments in that scene (FHV$_{POFL}$). The FHV$_{PPFL}$, while easy to implement on current hardware, exhibits a directional bias of the fragments with respect to the original view setup at the time of rasterization, which leads to artifacts. Additionally, FHV$_{PPFL}$ also incurs an overhead when creating images for novel viewpoints by excessive re-processing of fragments. In contrast, FHV$_{POFL}$ using per-octant lists of fragments provides a view-independent representation of the scene, which allows for ray-based evaluation and produces non-biased fragments as well as an optimal amount of fragments by using a triangle's tangent-space transform. FHV$_{POFA}$, using per-octant arrays of fragments, provides the same advantages as FHV$_{POFL}$ as well as the ability of reducing memory requirements using a two-pass initialization technique. Additionally, FHV$_{POFA}$ stores spatially close fragments together in GPU memory potentially reducing cache misses when per-octant fragments are accessed.

Similar to deferred shading FHVs trade computational effort for bandwidth. However, where deferred shading minimizes the amount of fragments to be processed, FHVs maximize the number of fragments. This provides the appearance-evaluation stage with additional opportunities to correctly handle transparency or incorporate shadow computation as well as employing indirect illumination techniques. Because FHVs are a discretized representation of a 3d scene, creating images for new viewpoints is essentially a resampling task. Visual appearance of those images is strongly connected to the spatial resolution of the octree representation and limited by the amount of physical memory on current graphics hardware. Performance results indicate potential for use in interactive real-time application scenarios.

## 6.1 Future Work

The technique presented here is obviously a first step. The appearance evaluation based on ray casting can be easily extended to handle indirect lighting and global illumination. However, this will certainly affect performance and needs to be developed in agreement with architectural constraints on current GPUs. In addition, the idea of an FHV could also be used to encode photon maps or represent light fields. Besides a further increase in memory requirements, such FHVs would also have to be updated after changes to light-source properties, which is not required for the technique presented here. Some octree-based global-illumination techniques (e.g., sparse voxel octrees [Crassin et al. 2009; Laine and Karras 2011], light propagation volumes [Kaplanyan and Dachsbacher 2010], irradiance volumes [Greger et al. 1998]), have problems with light leaking when voxels are sampled from both sides of a thin wall in a scene. Our octree-based FHV could be a solution to this because it stores a point cloud instead of a voxel volume at the leaf octants.

FHVs currently store fragment information without a geometric size, which means that the original rasterizer setup and the number of octree levels decide the spatial resolution. If the octree is exchanged for a bounding-volume hierarchy then only the rasterizer setup becomes the defining factor. In addition, fragments then could be interpreted as triangles when evaluating appearance for images from novel viewpoints. Another possible direction would be to invert the geometry representation by storing primitive-object information (e.g., type, transformation) instead of fragments. This would alleviate the resampling problem for image generation but also necessitates a complete redesign of how spatial information is handled by the controlling process on the CPU, which seems to be a current trend in graphics API development for hardware-based rendering anyway.

We are considering a re-implementation using a compute API. This would provide many advantages in terms of flexibility as well as avoid many restrictions with respect to scene input and processing order imposed by the current graphics-pipeline definition. However, it would also require emulation of the rasterization stage, which will very likely not attain the performance of current hardware implementations [Laine and Karras 2011]. An alternative might be a hybrid technique exploiting the hardware rasterization stage of the graphics pipeline together with one or more compute passes using custom task-scheduling strategies [Steinberger et al. 2014].

Computer graphics includes a *duality* allowing image generation as an *object-order* or *image-order* process [Levoy 2007]. Hardware-based graphics is still based on object-order evaluation but there is a clear trend towards image-order techniques even for interactive real-time rendering. Our FHV technique may provide another (small) stepping stone to leverage the advantages of both of these (process) views.